\documentclass[aip, reprint, amsmath, amssymb]{revtex4-1}

\usepackage{graphicx}
\usepackage{dcolumn}
\usepackage{bm}
\usepackage{float}
\usepackage{dblfloatfix}
\usepackage[french]{babel}
\usepackage[utf8]{inputenc}
\usepackage[T1]{fontenc}
\begin{document}

\title{Nonequilibrium Fluctuations and Intrinsic Stability of Nanostructured Solar Cells: Microscopic Derivation and Universal Laws}

\author{Rivo Herivola Manjakamanana Ravelonjato$^{*}$}
\affiliation{INSTN, Antananarivo, Madagascar}

\author{Ravo Tokiniaina Ranaivoson}
\affiliation{INSTN, Antananarivo, Madagascar}

\author{Philipe Manjakasoa Randriantsoa}
\affiliation{INSTN, Antananarivo, Madagascar}

\author{Raoelina Andriambololona}
\affiliation{INSTN, Antananarivo, Madagascar}

\author{Wilfrid Chrysante Solofoarisina}
\affiliation{INSTN, Antananarivo, Madagascar}

\begin{abstract}
\noindent{\footnotesize $^{*}$Corresponding author: manjakamanana@yahoo.fr}

\medskip

We study nonequilibrium fluctuations of efficiency in nanostructured solar cells. Our contribution is threefold. First, we derive a microscopic stochastic model of efficiency from a two-state master equation describing absorption, radiative recombination, and extraction. This derivation establishes an explicit link between efficiency fluctuations and entropy production, and justifies the thermodynamic affinity through local detailed balance. Second, we review universal fluctuation laws established in the recent literature and apply them for the first time to the real parameters of nanostructured solar cells. We show that the total entropy production behaves as a submartingale, and that its negative exponential is a martingale. For a biased discrete random walk, the exact infimum law is verified numerically in the near-equilibrium regime, then applied to realistic solar cell configurations. Third, our simulations show that nanostructured solar cells operate in a strongly nonequilibrium regime, where negative entropy fluctuations, and hence efficiency fluctuations, are exponentially rare. Over all simulated trajectories, most configurations show no detectable negative fluctuation, consistent with the theoretical prediction. These results explain the intrinsic statistical stability of the devices and provide a conceptual and quantitative framework for understanding efficiency fluctuations around the mean values. They do not modify the mean efficiencies obtained previously, but clarify their statistical robustness in the strongly nonequilibrium regime.
\end{abstract}


\maketitle

\section{Introduction}

The operation of solar cells is intrinsically a nonequilibrium process. The absorption of solar photons, the generation of charge carriers, and their extraction to an external circuit create a dissipative steady state, characterized by positive entropy production and significant fluctuations. Understanding these fluctuations is essential to assess the stability and reliability of devices, especially in nanostructures where the number of carriers is small and fluctuations become significant.

Since the pioneering work of Shockley and Queisser \cite{Shockley1961}, the efficiency limit of p-n junction cells has been well understood within the framework of detailed balance. For nanostructures, quantum confinement profoundly modifies the electronic properties \cite{Nozik2010,Carey2015}. Several attempts to adapt the Shockley-Queisser model to quantum dots have been reported \cite{Sahin2018,Alexandre2025,Anttu2015,Seo2025,Markvart2022,Ogundejj2026}. The best silicon cells plateau at 27.09\% \cite{Wang2024} and 27.0\% \cite{Xie2025}.

In a previous work \cite{Ravelonjato2026AIP}, we extended the Shockley-Queisser limit using the quantum phase space (QPS) formalism. This formalism was developed in \cite{Ranaivoson2022,Ravelonjato2023}, and validated for the thermodynamic properties of confined Fermi gases in \cite{Ravelonjato2026Nano}. The article \cite{Ravelonjato2026AIP} provided the mean efficiency values for different geometries (cube, parallelepiped, cylinder, sphere). These mean values are recalled in Appendix H, and their full derivations are reproduced in Appendices B–G of the present article.

We do not propose here a new extension of the Shockley-Queisser limit. The extension of this limit was carried out in our previous work \cite{Ravelonjato2026AIP}. The present article focuses on a complementary and distinct aspect: efficiency fluctuations around these mean values.

The present article has three objectives: (1) derive a microscopic stochastic model of efficiency from a two-state master equation, and establish the explicit link between efficiency fluctuations and entropy production; (2) justify the thermodynamic affinity $A$ through local detailed balance; (3) present a review of universal fluctuation laws established in the literature \cite{Neri2017,Neri2019,Barato2018,Doob1953,Roldan2024,Evans2002,Gallavotti1995} and apply them for the first time to the real parameters of nanostructured solar cells. We clarify what these laws bring conceptually, what they do not modify, and we identify an original result: solar cells operate in a strongly nonequilibrium regime ($A > 7$) where negative entropy fluctuations are exponentially rare.

Solar cells do not operate at thermodynamic equilibrium, but in a dissipative steady state. The classical Shockley-Queisser model \cite{Shockley1961} uses detailed balance as an approximation to calculate the radiative limit. This approximation is excellent for mean values, but it cannot describe fluctuations. The nonequilibrium analysis presented here does not replace the QPS model at effective equilibrium for calculating mean values; it adds a descriptive layer on fluctuations around these means.

The article is organized as follows. Section II briefly recalls the QPS formalism and the mean values calculated in \cite{Ravelonjato2026AIP}. Section III presents the microscopic derivation of the stochastic efficiency model and the justification of the affinity $A$. Section IV presents martingale theory and fluctuation laws. Section V presents the original numerical results. Section VI discusses the results and Section VII concludes. The derivation of the effective Hamiltonian from invariance under linear canonical transformations is reproduced in Appendix A. The full derivations of the QPS model (Appendices A–F of \cite{Ravelonjato2026AIP}) are reproduced in Appendices B–G. The tables of mean values are in Appendix H. The intermediate calculations of the stochastic model are in Appendices I, J, and K.

\section{Recall of the QPS Formalism and Mean Values}
\label{sec:qps}

This section briefly recalls the QPS formalism and the equations of the SQ limit extension. The reader is referred to \cite{Ranaivoson2022,Ravelonjato2023,Ravelonjato2026Nano,Ravelonjato2026AIP} for details. The complete derivation of the effective Hamiltonian is reproduced in Appendix A, and the complementary derivations of the QPS model are reproduced in Appendices B–G.

\subsection{QPS Formalism}
\label{sec:qps_formalism}

The quantum phase space (QPS) formalism relies on an effective Hamiltonian for a confined particle \cite{Ranaivoson2022,Ravelonjato2023}:
\begin{equation}
H = \sum_{l=1}^{3} \left(2\pi_l^\dagger \pi_l + 1\right) \frac{B_{ll}}{m^*},
\label{eq:hamiltonian}
\end{equation}
where $H$ is the effective Hamiltonian (in J), $\pi_l$ and $\pi_l^\dagger$ are bosonic ladder operators satisfying $[\pi_l, \pi_l^\dagger] = 1$, $B_{ll}$ is the momentum variance in direction $l$ (in kg$^2$ m$^2$/s$^2$), and $m^*$ is the effective carrier mass (in kg). The associated eigenenergies are $\varepsilon_{n_1,n_2,n_3} = \sum_l (2n_l+1) B_{ll}/m^*$ with $n_l = 0,1,2,\dots$.

— This Hamiltonian (\ref{eq:hamiltonian}) is obtained as a quadratic operator invariant under linear canonical transformations (LCT), which are linear transformations preserving the canonical commutation relations \cite{Ranaivoson2022}. The starting point is the harmonic oscillator Hamiltonian, which is invariant under the particular LCTs
\begin{equation}
\begin{pmatrix} p' \\ x' \end{pmatrix} = \begin{pmatrix} \cos\theta & -m\omega\sin\theta \\ \frac{\sin\theta}{m\omega} & \cos\theta \end{pmatrix} \begin{pmatrix} p \\ x \end{pmatrix},
\label{eq:lct_example}
\end{equation}
where $p'$ and $x'$ are the transformed momentum and position operators (in kg$\cdot$m/s and m respectively), $p$ and $x$ are the original momentum and position operators (in kg$\cdot$m/s and m), $\theta$ is a real dimensionless parameter (expressed in radians), $m$ is the particle mass (in kg), and $\omega$ is the oscillator angular frequency (in rad/s). This transformation preserves the canonical commutation relation $[p', x'] = [p, x] = -i\hbar$, where $\hbar$ is the reduced Planck constant (in J$\cdot$s). By rewriting this Hamiltonian in terms of the momentum variance $B_{ll}$ and generalizing to a free particle confined in three directions, we obtain the form (\ref{eq:hamiltonian}) \cite{Ranaivoson2022,Ravelonjato2023}. The complete derivation, step by step, is reproduced in \textbf{Appendix A}. The Hamiltonian (\ref{eq:hamiltonian}) is therefore an \emph{effective construction} constrained by two principles: (i) invariance under LCT, and (ii) reproduction of the classical and degenerate quantum limits (Appendix A).

The momentum variance $B_{ll}$ is obtained from the discrete and classical grand canonical potential in the classical limit (B1-B4). For each geometry, $B_{ll} = (\hbar / 2L_l) \Delta$, where $L_l$ is the effective length in direction $l$ (in m). The product $\prod_l B_{ll} = (\hbar/2)^3 (\Delta^3/V)$ is independent of the individual dimensions at constant volume $V$. The validity of this unified expression has been confirmed by its application to confined Fermi gases \cite{Ravelonjato2026Nano}, and the complete derivation is reproduced in \textbf{Appendix B}.

The universal parameter $\Delta$ is \cite{Ravelonjato2023,Ravelonjato2026Nano}:
\begin{equation}
\Delta = \sqrt{2\pi m^* k_B T + \pi \hbar^2 \left(6\pi^2 \frac{n}{g_s}\right)^{2/3} \frac{m}{m^*} + \theta m \Delta\epsilon_{\text{int}}},
\label{eq:delta}
\end{equation}
where $k_B$ is the Boltzmann constant ($1.381 \times 10^{-23}$ J/K), $T$ is the absolute temperature (in K), $\hbar$ is the reduced Planck constant ($1.055 \times 10^{-34}$ J$\cdot$s), $n$ is the carrier density (in m$^{-3}$), $g_s$ is the spin degeneracy factor ($g_s = 2$ for electrons), $m$ is the reduced effective mass (geometric mean of $m_e$ and $m_h$), $\theta$ is an adjustable coefficient ($\theta = 0$ for an ideal gas, $\theta = 1$ with interactions), and $\Delta\epsilon_{\text{int}}$ is the exchange-correlation correction (in eV).

\subsection{Mean Efficiency Values}

The extension of the SQ limit for nanostructures was recently developed in \cite{Ravelonjato2026AIP}. The main equations are:
\begin{equation}
J_0^{\text{QD}} = q\,2A_p\,t_e\,Q(E_g^{\text{eff}},T_c)\; \frac{\lambda_{\text{th}}^3 m^3}{8\beta^3 (\hbar/2)^3 \Delta^3},
\label{eq:j0qd}
\end{equation}
where $J_0^{\text{QD}}$ is the radiative saturation current (in A/m$^2$), $q$ is the elementary charge ($1.602 \times 10^{-19}$ C), $A_p$ is the projected cell area (in m$^2$), $t_e$ is the absorption efficiency ($t_e = 1$), $Q(E_g^{\text{eff}},T_c)$ is the photon flux emitted by the cell at temperature $T_c$ (in photons$\cdot$s$^{-1}$$\cdot$m$^{-2}$), $E_g^{\text{eff}}$ is the effective bandgap (in eV), $T_c$ is the cell temperature (in K), $\lambda_{\text{th}} = h/\sqrt{2\pi m k_B T_c}$ is the thermal de Broglie wavelength (in m), $h$ is the Planck constant ($6.626 \times 10^{-34}$ J$\cdot$s), $m$ is the reduced effective mass (in kg), and $\beta = 1/(k_B T_c)$ (in J$^{-1}$). The complete derivation of equation (\ref{eq:j0qd}) is reproduced in Appendix C.

\begin{equation}
J_{sc} = q\,A_p\,t_s\,f_\omega\,Q_s(E_g^{\text{eff}}),
\label{eq:jsc}
\end{equation}
where $J_{sc}$ is the short-circuit current (in A/m$^2$), $t_s$ is the collection efficiency ($t_s = 1$), $f_\omega = \omega_s/\pi \approx 2.18 \times 10^{-5}$ is the geometric factor related to the sun's solid angle, and $Q_s(E_g^{\text{eff}})$ is the solar photon flux (blackbody at $T_s = 6000$ K, in photons$\cdot$s$^{-1}$$\cdot$m$^{-2}$).

\begin{equation}
V_{oc} = \frac{k_B T_c}{q} \ln\!\left(\frac{J_{sc}}{J_0}\right),
\label{eq:voc}
\end{equation}
where $V_{oc}$ is the open-circuit voltage (in V), $J_{sc}$ is the short-circuit current defined in (\ref{eq:jsc}), $J_0 = J_0^{\text{QD}}$ is the radiative saturation current defined in (\ref{eq:j0qd}), $k_B$ is the Boltzmann constant, $T_c$ is the cell temperature, and $q$ is the elementary charge.

\begin{equation}
FF = \frac{v_{oc} - \ln(v_{oc}+0.72)}{v_{oc}+1}, \quad v_{oc} = \frac{qV_{oc}}{k_B T_c},
\label{eq:ff}
\end{equation}
where $FF$ is the fill factor (dimensionless), $v_{oc}$ is the normalized open-circuit voltage (dimensionless), $V_{oc}$ is the open-circuit voltage defined in (\ref{eq:voc}), and $q$, $k_B$, $T_c$ are defined previously.

\begin{equation}
\eta = \frac{J_{sc}\,V_{oc}\,FF}{P_{\text{inc}}}, \quad P_{\text{inc}} = 100\,\mathrm{mW/cm^2},
\label{eq:eta}
\end{equation}
where $\eta$ is the conversion efficiency (in \%), $J_{sc}$ is defined in (\ref{eq:jsc}), $V_{oc}$ in (\ref{eq:voc}), $FF$ in (\ref{eq:ff}), and $P_{\text{inc}}$ is the incident power (in W/m$^2$).

The confinement energies $E_g^{\text{eff}}$ for the four geometries are given in \cite{Ravelonjato2026AIP} and reproduced in Appendix B. The resulting mean efficiency values are recalled in Appendix H. For reference, the 5 nm cube reaches $\eta = 48.7$ \%, the 3 nm sphere reaches $\eta = 49.1$ \%, the cylinder of height 3.5 nm and the parallelepiped of height 3.5 nm reach $\eta = 49.0$ \%. The numerical method used is described in Appendix D, and the verification of the return to classical values for $L \to \infty$ is given in Appendix E.

These mean values \cite{Ravelonjato2026AIP} serve as our reference for the fluctuation analysis.

\section{Stochastic Model of Efficiency}
\label{sec:modele}

This section presents the microscopic derivation of the stochastic efficiency model. The detailed intermediate calculations are grouped in Appendices I, J, and K.

\subsection{Microscopic Processes and Master Equation}
\label{sec:master}

We model a nanostructured solar cell as a two-state system:
\begin{itemize}
\item $|0\rangle$: ground state (no exciton);
\item $|1\rangle$: excited state (one electron-hole pair).
\end{itemize}

Three elementary processes are considered:
\begin{enumerate}
\item \textbf{Absorption}: $|0\rangle \to |1\rangle$, rate $k_{\mathrm{abs}}$, from the solar photon reservoir at $T_s = 6000$ K.
\item \textbf{Radiative recombination}: $|1\rangle \to |0\rangle$, rate $k_{\mathrm{rad}}$, emitting a photon into the thermal bath at $T_c = 300$ K.
\item \textbf{Extraction}: $|1\rangle \to |0\rangle$, rate $k_{\mathrm{ext}}$, transferring an electron to the external circuit at potential $V$.
\end{enumerate}

The master equation for the probabilities $P_0(t)$ and $P_1(t)$ reads (Appendix I):
\begin{align}
\frac{dP_1}{dt} &= k_{\mathrm{abs}} P_0 - (k_{\mathrm{rad}} + k_{\mathrm{ext}}) P_1, \label{eq:master1}\\
\frac{dP_0}{dt} &= -k_{\mathrm{abs}} P_0 + (k_{\mathrm{rad}} + k_{\mathrm{ext}}) P_1, \label{eq:master0}
\end{align}
where $P_0(t)$ and $P_1(t)$ are the occupation probabilities of states $|0\rangle$ and $|1\rangle$ (dimensionless, with $P_0 + P_1 = 1$), $k_{\mathrm{abs}}$ is the absorption rate (in s$^{-1}$), $k_{\mathrm{rad}}$ is the radiative recombination rate (in s$^{-1}$), and $k_{\mathrm{ext}}$ is the extraction rate (in s$^{-1}$).

In the steady state, we obtain:
\begin{equation}
P_1^{\mathrm{ss}} = \frac{k_{\mathrm{abs}}}{k_{\mathrm{abs}} + k_{\mathrm{rad}} + k_{\mathrm{ext}}}, \quad
P_0^{\mathrm{ss}} = \frac{k_{\mathrm{rad}} + k_{\mathrm{ext}}}{k_{\mathrm{abs}} + k_{\mathrm{rad}} + k_{\mathrm{ext}}},
\label{eq:stationary}
\end{equation}
where $P_0^{\mathrm{ss}}$ and $P_1^{\mathrm{ss}}$ are the steady-state probabilities of states $|0\rangle$ and $|1\rangle$ (dimensionless), and $k_{\mathrm{abs}}$, $k_{\mathrm{rad}}$, $k_{\mathrm{ext}}$ are the rates defined in (\ref{eq:master1}).

\subsection{Stochastic Efficiency}
\label{sec:efficacite_stoch}

Over a trajectory of duration $t$, we count:
\begin{itemize}
\item $N_{\mathrm{abs}}(t)$: number of absorptions (dimensionless);
\item $N_{\mathrm{rad}}(t)$: number of radiative recombinations (dimensionless);
\item $N_{\mathrm{ext}}(t)$: number of extractions (dimensionless).
\end{itemize}

The energy balance per trajectory reads:
\begin{equation}
N_{\mathrm{abs}} E_g^{\mathrm{eff}} = N_{\mathrm{rad}} E_g^{\mathrm{eff}} + N_{\mathrm{ext}} qV + Q_{\mathrm{cell}},
\label{eq:bilan_energetique}
\end{equation}
where $N_{\mathrm{abs}}$, $N_{\mathrm{rad}}$, $N_{\mathrm{ext}}$ are the event counts defined above, $E_g^{\mathrm{eff}}$ is the effective bandgap (in eV), $q$ is the elementary charge (in C), $V$ is the operating voltage (in V), and $Q_{\mathrm{cell}}$ is the heat dissipated in the cell (in eV).

Using the conservation $N_{\mathrm{abs}} = N_{\mathrm{rad}} + N_{\mathrm{ext}}$ (each absorption leads either to a recombination or an extraction), we obtain:
\begin{equation}
Q_{\mathrm{cell}} = N_{\mathrm{ext}} (E_g^{\mathrm{eff}} - qV),
\label{eq:qcell}
\end{equation}
where $Q_{\mathrm{cell}}$ is the dissipated heat (in eV), $N_{\mathrm{ext}}$ is the number of extractions, $E_g^{\mathrm{eff}}$ is the effective bandgap (in eV), $q$ is the elementary charge, and $V$ is the operating voltage.

The instantaneous stochastic efficiency is defined as the ratio of useful extracted energy to absorbed solar energy:
\begin{equation}
\eta(t) = \frac{qV \cdot N_{\mathrm{ext}}(t)}{E_g^{\mathrm{eff}} \cdot N_{\mathrm{abs}}(t)},
\label{eq:eta_stoch}
\end{equation}
where $\eta(t)$ is the instantaneous stochastic efficiency (dimensionless), $q$ is the elementary charge (in C), $V$ is the operating voltage (in V), $N_{\mathrm{ext}}(t)$ is the number of extractions up to time $t$ (dimensionless), $E_g^{\mathrm{eff}}$ is the effective bandgap (in eV), and $N_{\mathrm{abs}}(t)$ is the number of absorptions up to time $t$ (dimensionless).

For long times, $N_{\mathrm{abs}}(t) \approx k_{\mathrm{abs}} t$ (law of large numbers), and the mean efficiency becomes:
\begin{equation}
\langle \eta \rangle = \frac{qV}{E_g^{\mathrm{eff}}} \cdot \frac{k_{\mathrm{ext}}}{k_{\mathrm{rad}} + k_{\mathrm{ext}}},
\label{eq:eta_moyenne}
\end{equation}
where $\langle \eta \rangle$ is the mean efficiency (dimensionless), $q$ is the elementary charge, $V$ is the operating voltage, $E_g^{\mathrm{eff}}$ is the effective bandgap, and $k_{\mathrm{ext}}$, $k_{\mathrm{rad}}$ are the rates defined in (\ref{eq:master1}).

\subsection{Entropy Production and Affinity}
\label{sec:entropie_affinite}

The total entropy production in the environment along a trajectory is (Appendix J):
\begin{equation}
S_{\mathrm{tot}}(t) = \Delta S_{\mathrm{sun}} + \Delta S_{\mathrm{cell}} + \Delta S_{\mathrm{circuit}},
\label{eq:stot_decomp}
\end{equation}
where $S_{\mathrm{tot}}(t)$ is the total entropy production (in units of $k_B$, dimensionless), $\Delta S_{\mathrm{sun}}$ is the entropy change of the solar reservoir, $\Delta S_{\mathrm{cell}}$ is the entropy change of the cell, and $\Delta S_{\mathrm{circuit}}$ is the entropy change of the external circuit (all in units of $k_B$), with
\begin{align}
\Delta S_{\mathrm{sun}} &= -\frac{N_{\mathrm{abs}} E_g^{\mathrm{eff}}}{k_B T_s}, \label{eq:sun}\\
\Delta S_{\mathrm{cell}} &= \frac{Q_{\mathrm{cell}}}{k_B T_c} = \frac{N_{\mathrm{ext}} (E_g^{\mathrm{eff}} - qV)}{k_B T_c}, \label{eq:cell}\\
\Delta S_{\mathrm{circuit}} &= 0 \quad \text{(work reservoir)}. \label{eq:circuit}
\end{align}
where $N_{\mathrm{abs}}$ is the number of absorptions, $E_g^{\mathrm{eff}}$ is the effective bandgap (in eV), $k_B$ is the Boltzmann constant (in J/K), $T_s = 6000$ K is the sun temperature, $Q_{\mathrm{cell}}$ is the dissipated heat defined in (\ref{eq:qcell}), $T_c = 300$ K is the cell temperature, $N_{\mathrm{ext}}$ is the number of extractions, $q$ is the elementary charge, and $V$ is the operating voltage.

Neglecting the sun contribution (limit $T_s \to \infty$, justified in Appendix K), we obtain:
\begin{equation}
S_{\mathrm{tot}}(t) = A \cdot N_{\mathrm{ext}}(t),
\label{eq:stot_A}
\end{equation}
where $S_{\mathrm{tot}}(t)$ is the total entropy production (dimensionless), $A$ is the thermodynamic affinity (dimensionless, defined below), and $N_{\mathrm{ext}}(t)$ is the number of extractions up to time $t$.

The thermodynamic affinity $A$ is defined by:
\begin{equation}
A = \frac{E_g^{\mathrm{eff}} - qV}{k_B T_c}.
\label{eq:affinite}
\end{equation}
where $A$ is the thermodynamic affinity (dimensionless), $E_g^{\mathrm{eff}}$ is the effective bandgap (in eV), $q$ is the elementary charge (in C), $V$ is the operating voltage (in V), $k_B$ is the Boltzmann constant (in J/K), and $T_c = 300$ K is the cell temperature.

\textbf{Physical interpretation}: $A$ measures the number of thermal energy units $k_B T_c$ dissipated per extracted carrier. It is the entropy production per extraction, in units of $k_B$.

\subsection{Justification of Detailed Balance and Affinity}
\label{sec:bilan_detaille}

The local detailed balance relation for the two de-excitation channels $|1\rangle \to |0\rangle$ reads (Appendix J):
\begin{equation}
\frac{k_{\mathrm{ext}}}{k_{\mathrm{rad}}} = \exp\left(\frac{E_g^{\mathrm{eff}} - qV}{k_B T_c}\right) = e^{A}.
\label{eq:bilan_detaille}
\end{equation}
where $k_{\mathrm{ext}}$ is the extraction rate (in s$^{-1}$), $k_{\mathrm{rad}}$ is the radiative recombination rate (in s$^{-1}$), $E_g^{\mathrm{eff}}$ is the effective bandgap (in eV), $q$ is the elementary charge, $V$ is the operating voltage, $k_B$ is the Boltzmann constant, $T_c$ is the cell temperature, and $A$ is the affinity defined in (\ref{eq:affinite}).

This relation justifies \emph{a posteriori} the biased random walk model used in the literature \cite{Neri2017}: the forward (extraction) and backward (recombination) hopping rates are in a ratio $e^{A}$.

\textbf{Microscopic justification of $A$}: local detailed balance (Appendix J) states that for a transition $i \to j$ via a channel $\nu$, the ratio of forward and backward rates is given by the exponential of the entropy change of the reservoirs:
\begin{equation}
\frac{k_{i \to j}^{\nu}}{k_{j \to i}^{\nu}} = \exp\left(\frac{\Delta S_{\mathrm{env}}^{\nu}}{k_B}\right).
\label{eq:ldb}
\end{equation}
where $k_{i \to j}^{\nu}$ is the transition rate from $i$ to $j$ via channel $\nu$ (in s$^{-1}$), $k_{j \to i}^{\nu}$ is the reverse transition rate (in s$^{-1}$), $\Delta S_{\mathrm{env}}^{\nu}$ is the entropy change of the environment for this transition (in J/K), and $k_B$ is the Boltzmann constant.

For the extraction channel, $\Delta S_{\mathrm{env}}^{\mathrm{ext}} = (E_g^{\mathrm{eff}} - qV)/T_c = k_B A$. For the radiative channel, $\Delta S_{\mathrm{env}}^{\mathrm{rad}} = E_g^{\mathrm{eff}}/T_c$. Comparing the two channels, we obtain relation (\ref{eq:bilan_detaille}).

\subsection{Reduction to a Biased Random Walk}
\label{sec:marche_aleatoire}

Combining equations (\ref{eq:stot_A}) and (\ref{eq:eta_stoch}), we establish the explicit link between efficiency fluctuations and entropy production:
\begin{equation}
\eta(t) = \frac{qV}{E_g^{\mathrm{eff}}} \cdot \frac{S_{\mathrm{tot}}(t)/A}{N_{\mathrm{abs}}(t)}.
\label{eq:lien_eta_S}
\end{equation}
where $\eta(t)$ is the instantaneous stochastic efficiency (dimensionless), $q$ is the elementary charge, $V$ is the operating voltage, $E_g^{\mathrm{eff}}$ is the effective bandgap, $S_{\mathrm{tot}}(t)$ is the total entropy production defined in (\ref{eq:stot_A}), $A$ is the affinity defined in (\ref{eq:affinite}), and $N_{\mathrm{abs}}(t)$ is the number of absorptions up to time $t$.

For long times, $N_{\mathrm{abs}}(t) \approx k_{\mathrm{abs}} t$ is quasi-deterministic (Poisson process), and the relative efficiency fluctuations equal the relative entropy fluctuations:
\begin{equation}
\frac{\delta \eta}{\langle \eta \rangle} = \frac{\delta S_{\mathrm{tot}}}{\langle S_{\mathrm{tot}} \rangle}.
\label{eq:fluct_rel}
\end{equation}
where $\delta \eta = \eta - \langle \eta \rangle$ is the efficiency fluctuation (dimensionless), $\langle \eta \rangle$ is the mean efficiency defined in (\ref{eq:eta_moyenne}), $\delta S_{\mathrm{tot}} = S_{\mathrm{tot}} - \langle S_{\mathrm{tot}} \rangle$ is the entropy production fluctuation (dimensionless), and $\langle S_{\mathrm{tot}} \rangle$ is its mean.

Thus, the probability of a negative efficiency fluctuation is controlled by the probability of a negative fluctuation of $S_{\mathrm{tot}}$, which is exponentially rare for large $A$ (Section IV).

The model reduces to a discrete random walk with:
\begin{itemize}
\item forward step (extraction): $S \to S + A$, rate $k_+ = k_{\mathrm{ext}} P_1^{\mathrm{ss}}$;
\item backward step (recombination): $S \to S - A$, rate $k_- = k_{\mathrm{rad}} P_1^{\mathrm{ss}}$.
\end{itemize}
where $S$ is the current entropy production (dimensionless), $A$ is the affinity defined in (\ref{eq:affinite}), $k_+$ and $k_-$ are the forward and backward hopping rates (in s$^{-1}$), $k_{\mathrm{ext}}$ and $k_{\mathrm{rad}}$ are defined in (\ref{eq:master1}), and $P_1^{\mathrm{ss}}$ is the steady-state probability defined in (\ref{eq:stationary}).

With $k_+/k_- = e^{A}$, we recover exactly the model of \cite{Neri2017}. The total entropy production reads:
\begin{equation}
S_{\mathrm{tot}}(t) = A \times (N_+(t) - N_-(t)),
\label{eq:stot_walk}
\end{equation}
where $S_{\mathrm{tot}}(t)$ is the total entropy production (dimensionless), $A$ is the affinity defined in (\ref{eq:affinite}), and $N_+(t)$, $N_-(t)$ are the numbers of forward and backward hops up to time $t$ (dimensionless).

\section{Martingale Theory and Fluctuation Laws}
\label{sec:martingales}

This section presents the mathematical tools used to derive the fluctuation laws. The reader is referred to \cite{Doob1953,Roldan2024} for the foundations.

\subsection{Stochastic Entropy Production}

For a mesoscopic system in contact with thermal reservoirs, the total entropy production $S_{\mathrm{tot}}(t)$ along a trajectory is \cite{Seifert2012}:
\begin{equation}
S_{\mathrm{tot}}(t) = \Delta S_{\mathrm{sys}}(t) + S_{\mathrm{env}}(t).
\label{eq:entropy}
\end{equation}
where $S_{\mathrm{tot}}(t)$ is the total entropy production (dimensionless), $\Delta S_{\mathrm{sys}}(t)$ is the system entropy change (dimensionless), and $S_{\mathrm{env}}(t)$ is the environment entropy change (dimensionless).

In a nonequilibrium steady state, it is expressed as the logarithm of the forward/backward probability ratio \cite{Neri2017}:
\begin{equation}
S_{\mathrm{tot}}(t) = \log \left[ \frac{p(X_1,\ldots,X_t)}{p(X_t,\ldots,X_1)} \right].
\label{eq:entropy_path}
\end{equation}
where $X_1, \dots, X_t$ denote the successive states of the trajectory in the interval $[0,t]$, $p(X_1,\ldots,X_t)$ is the forward trajectory probability, and $p(X_t,\ldots,X_1)$ is the backward (time-reversed) trajectory probability.

\subsection{Negative Exponential Martingale}

Fundamental property: $\exp(-S_{\mathrm{tot}}(t))$ is a martingale \cite{Neri2017,Neri2019,Roldan2024,Doob1953}:
\begin{equation}
\langle \exp(-S_{\mathrm{tot}}(t)) | \mathcal{F}_s \rangle = \exp(-S_{\mathrm{tot}}(s)), \quad \forall s \leq t.
\label{eq:martingale}
\end{equation}
where $\langle \cdot | \mathcal{F}_s \rangle$ denotes the conditional expectation with respect to the filtration $\mathcal{F}_s$ generated by the process up to time $s$, and $s \leq t$ means the property holds for any later time.

Hence the integral fluctuation relation:
\begin{equation}
\langle \exp(-S_{\mathrm{tot}}(t)) \rangle = 1.
\label{eq:integral_fluctuation}
\end{equation}
where $\langle \cdot \rangle$ denotes the expectation (mean) over all possible trajectories.

\subsection{Second Law at Stopping Times}

By Jensen, we obtain $\langle S_{\mathrm{tot}}(t) \rangle \geq 0$. More generally, $S_{\mathrm{tot}}$ is a submartingale:
\begin{equation}
\langle S_{\mathrm{tot}}(t) | \mathcal{F}_s \rangle \geq S_{\mathrm{tot}}(s).
\label{eq:submartingale}
\end{equation}
where the inequality (instead of equality for a martingale) means that entropy production has a positive drift on average.

Using Doob's optional stopping theorem \cite{Doob1953}, for a stopping time $T$:
\begin{equation}
\langle \exp(-S_{\mathrm{tot}}(T)) \rangle = 1, \quad \langle S_{\mathrm{tot}}(T) \rangle \geq 0.
\label{eq:stopping}
\end{equation}
where $T$ is a stopping time, i.e., a random time defined by a causal criterion (depending only on the past of the trajectory). This second law at stopping times means that it is not possible to reduce entropy on average, even by stopping the process at a cleverly chosen moment.

\subsection{Universal Laws for Extremes}

The global infimum $S_{\mathrm{inf}} = \inf_{t \geq 0} S_{\mathrm{tot}}(t)$ follows an exponential law in the continuous limit \cite{Neri2017}:
\begin{equation}
p_{S_{\mathrm{inf}}}(s) = e^s, \quad s \leq 0,
\label{eq:infimum_distribution}
\end{equation}
where $S_{\mathrm{inf}}$ is the minimum value (infimum) of entropy production over the entire trajectory, and $p_{S_{\mathrm{inf}}}(s)$ is its probability density for $s \leq 0$ (the infimum can only take negative or zero values). This distribution implies a mean $\langle S_{\mathrm{inf}} \rangle = -1$. Hence the bound:
\begin{equation}
\mathbb{P}(S_{\mathrm{inf}} \leq -s) \leq e^{-s}, \quad s \geq 0,
\label{eq:infimum_bound}
\end{equation}
where $\mathbb{P}(\cdot)$ denotes the probability of an event, and $s \geq 0$ is the threshold considered.

For a discrete random walk with affinity $A$, the exact formula is \cite{Neri2017}:
\begin{equation}
\langle S_{\mathrm{inf}} \rangle = -\frac{A}{e^A - 1}.
\label{eq:infimum_exact}
\end{equation}
where $\langle S_{\mathrm{inf}} \rangle$ is the mean infimum of entropy production (dimensionless), $A$ is the thermodynamic affinity defined in (\ref{eq:affinite}) (dimensionless), and $e$ is the base of the natural logarithm.

\subsection{Passage Probabilities and Stopping Times}

For a first passage time $T_{\mathrm{fp}}$ out of $(-s_-, s_+)$, the probabilities $p_+$ and $p_-$ are \cite{Neri2017}:
\begin{equation}
p_+ = \frac{e^{s_-} - 1}{e^{s_-} - e^{-s_+}}, \quad p_- = \frac{1 - e^{-s_+}}{e^{s_-} - e^{-s_+}}.
\label{eq:splitting}
\end{equation}
where $s_+ > 0$ and $s_- > 0$ are the exit thresholds (asymmetric in general), $p_+$ is the probability of exiting through the positive threshold, and $p_-$ is the probability of exiting through the negative threshold.

For symmetric first passage times $T_+$ and $T_-$:
\begin{equation}
\frac{p_{T_+}(t; s)}{p_{T_-}(t; -s)} = e^s, \quad \langle T_+ \rangle = \langle T_- \rangle,
\label{eq:stopping_fluctuation}
\end{equation}
where $p_{T_+}(t; s)$ is the probability density of the first passage time through the positive threshold $s$, $p_{T_-}(t; -s)$ is the probability density of the first passage time through the negative threshold $-s$, and $\langle T_+ \rangle$, $\langle T_- \rangle$ are the corresponding mean first passage times.

\subsection{Arcsine Laws}

For a thermodynamic current $X(t)$, the fraction of time $T_1$ spent above its mean follows the arcsine law \cite{Barato2018}:
\begin{equation}
P(T_1) = \frac{1}{\pi} \frac{1}{\sqrt{T_1(1 - T_1)}},
\label{eq:arcsine}
\end{equation}
where $T_1 \in [0,1]$ is the fraction of time spent above the mean, and $P(T_1)$ is its probability density.

\subsection{Fluctuation Theorem}

The fluctuation theorem \cite{Evans2002,Gallavotti1995}:
\begin{equation}
\frac{p(S_{\mathrm{tot}} = A)}{p(S_{\mathrm{tot}} = -A)} = e^{A},
\label{eq:fluctuation_theorem}
\end{equation}
where $p(S_{\mathrm{tot}} = A)$ is the probability of observing an entropy production equal to $A$, and $p(S_{\mathrm{tot}} = -A)$ is the probability of observing the opposite value $-A$. This exponential ratio quantifies the asymmetry between positive and negative fluctuations.

\section{Numerical Results}
\label{sec:results}

This section presents the original numerical results. We verify the exact infimum law formula for $A < 2$, then apply it to the real parameters of nanostructured solar cells.

\subsection{Model and Simulation Method}

We model entropy production as a biased random walk with affinity $A$, derived microscopically in Section III. The affinity is given by equation (\ref{eq:affinite}). The hopping rates are:
\begin{equation}
k_\pm = k_0 \exp(\pm A/2),
\label{eq:rates}
\end{equation}
where $k_+$ and $k_-$ are the forward and backward hopping rates (in s$^{-1}$), $k_0$ is a reference rate ($k_0 = 1$ in reduced units), and $A$ is the affinity defined in (\ref{eq:affinite}).

The simulations use the Gillespie algorithm (continuous time). For $A < 2$, we use $n_{\text{traj}} = 10^4$ trajectories and $t_{\text{max}} = 500$. For $A > 7$, we use $n_{\text{traj}} = 500$ trajectories and $t_{\text{max}} = 5$.

\subsection{Numerical Verification of the Infimum Law}

Table \ref{tab:verification} compares the simulations with formula (\ref{eq:infimum_exact}). The agreement is excellent: all deviations are less than $2\sigma$.

\begin{table}[H]
\centering
\caption{Verification of the formula $\langle S_{\mathrm{inf}} \rangle = -A/(e^A - 1)$ \cite{Neri2017}. Simulations with $n_{\text{traj}} = 10^4$ and $t_{\text{max}} = 500$. The “Deviation” column gives the difference between simulation and theory.}
\label{tab:verification}
\begin{ruledtabular}
\begin{tabular}{cccc}
$A$ & Simulation & Theory \cite{Neri2017} & Deviation \\
\hline
0.2 & $-0.9176 \pm 0.0103$ & $-0.9033$ & $-0.0143$ \\
0.5 & $-0.7762 \pm 0.0100$ & $-0.7707$ & $-0.0054$ \\
1.0 & $-0.5887 \pm 0.0096$ & $-0.5820$ & $-0.0067$ \\
1.5 & $-0.4161 \pm 0.0090$ & $-0.4308$ & $0.0147$ \\
2.0 & $-0.3208 \pm 0.0088$ & $-0.3130$ & $-0.0078$ \\
\end{tabular}
\end{ruledtabular}
\end{table}

\begin{figure}[H]
\centering
\includegraphics[width=0.48\textwidth]{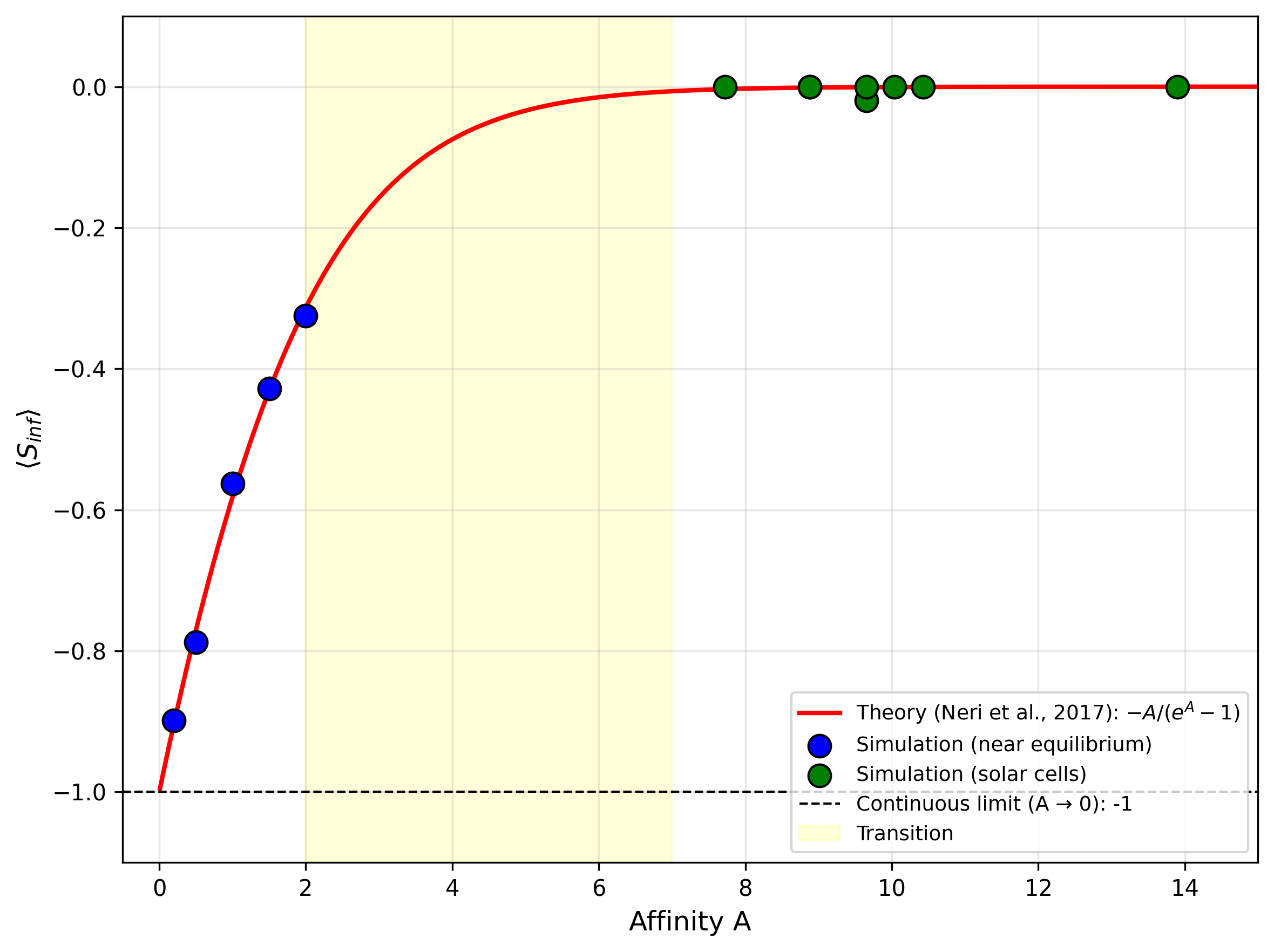}
\caption{Infimum law: exact formula and simulations. Blue points ($A < 2$) follow the theoretical curve $-A/(e^A - 1)$ (red line) \cite{Neri2017}. Green points (solar cells, $A > 7$) are close to 0. The black dashed line at $-1$ is the continuous limit ($A \to 0$). The yellow region marks the transition ($2 < A < 7$).}
\label{fig:infimum}
\end{figure}

Figure \ref{fig:infimum} shows $\langle S_{\mathrm{inf}} \rangle$ as a function of $A$.

Three distinct regimes appear:
\begin{itemize}
\item \textbf{Near-equilibrium regime} ($A < 2$): agreement with the theoretical curve.
\item \textbf{Transition regime} ($2 < A < 7$): $\langle S_{\mathrm{inf}} \rangle$ deviates from $-1$.
\item \textbf{Strongly nonequilibrium regime} ($A > 7$): $\langle S_{\mathrm{inf}} \rangle \approx 0$, negative fluctuations exponentially rare.
\end{itemize}

\begin{figure}[H]
\centering
\includegraphics[width=0.48\textwidth]{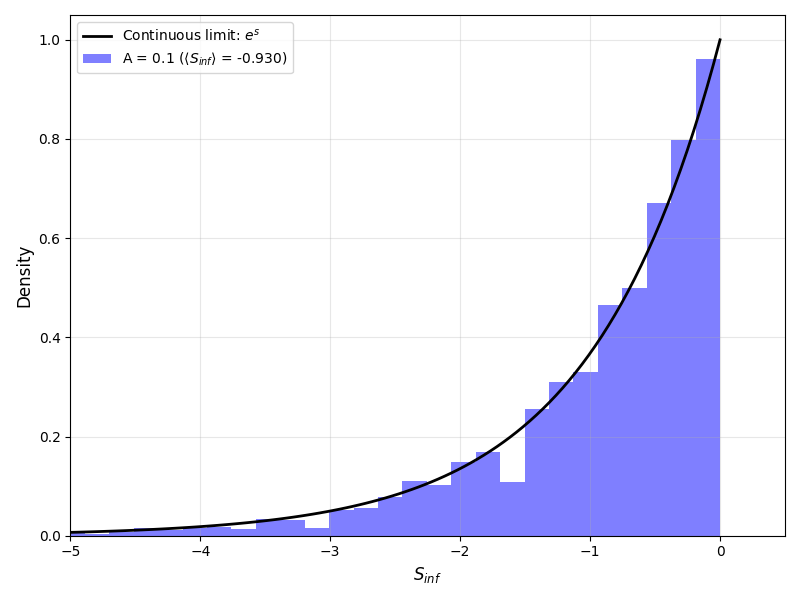}
\caption{Convergence to the continuous law ($A \to 0$). Distribution of $S_{\mathrm{inf}}$ for $A = 0.1$ compared with the continuous law $e^s$ \cite{Neri2017}. The simulation gives $\langle S_{\mathrm{inf}} \rangle = -0.9316$, close to theory ($-0.9508$) and to the continuous limit $-1$.}
\label{fig:convergence}
\end{figure}

Figure \ref{fig:convergence} shows the convergence to the continuous law for $A = 0.1$.

\subsection{Application to Nanostructured Solar Cells}

The affinities for the configurations in Tables H1-H4 range between 7.7 and 13.9. Formula (\ref{eq:infimum_exact}) gives the results in Table \ref{tab:solar_infimum}.

\begin{table}[H]
\centering
\caption{Mean infimum for solar cells. Simulations with $n_{\text{traj}} = 500$ and $t_{\text{max}} = 5$. The “Theory” column gives formula (\ref{eq:infimum_exact}) and the “Simulation” column gives the numerical results.}
\label{tab:solar_infimum}
\begin{ruledtabular}
\begin{tabular}{cccccc}
Configuration & $A$ & $\eta$ (\%) & $\langle S_{\mathrm{inf}} \rangle$ (theory) & $\langle S_{\mathrm{inf}} \rangle$ (simulation) \\
\hline
Cube 3 nm & 13.90 & 13.5 & $-1.28 \times 10^{-5}$ & $0.0000$ \\
Cube 5 nm & 10.42 & 48.7 & $-3.10 \times 10^{-4}$ & $0.0000$ \\
Cube 7 nm & 8.88 & 43.6 & $-1.24 \times 10^{-3}$ & $0.0000$ \\
Cube 10 nm & 7.72 & 32.2 & $-3.42 \times 10^{-3}$ & $0.0000$ \\
Sphere 3 nm & 10.04 & 49.1 & $-4.39 \times 10^{-4}$ & $-0.0201$ \\
Sphere 4 nm & 8.88 & 43.9 & $-1.24 \times 10^{-3}$ & $0.0000$ \\
Cylinder H=3.5 & 9.65 & 49.0 & $-6.20 \times 10^{-4}$ & $0.0000$ \\
Parallelepiped b=3.5 & 9.65 & 49.0 & $-6.20 \times 10^{-4}$ & $0.0000$ \\
\end{tabular}
\end{ruledtabular}
\end{table}

\textbf{Key observation}: for all high-efficiency configurations ($\eta > 40$ \%), $A > 8$, and the theoretical $\langle S_{\mathrm{inf}} \rangle$ is less than $10^{-3}$. With 500 trajectories, most configurations give $0.0000$, consistent with the predicted rarity ($e^{-A} \approx 10^{-4}$ to $10^{-6}$). For the 3 nm sphere, a negative fluctuation was observed ($-0.0201$), illustrating the existence of these rare events.

\subsection{Physical Consequence: Statistical Stability}

Using relation (\ref{eq:fluct_rel}), the probability of a negative efficiency fluctuation is of order $e^{-A}$, i.e., $e^{-10} \approx 5 \times 10^{-5}$ for the 5 nm cube. This means that the mean efficiency values are statistically robust in the regime $A > 7$.

\textbf{Important caveat}: this is a \emph{statistical} stability of entropy production and, by extension, of efficiency fluctuations. It does not prejudge the operational stability of devices (degradation, non-radiative recombination, contacts, etc.).

\subsection{Passage Probabilities}

The simulations confirm that $p_+ \approx 1$ for all configurations. The process almost always exits through the positive threshold. The probability of exiting through the negative threshold is exponentially small.

\section{Discussion}
\label{sec:discussion}

\subsection{Scope and Limitations of the Approach}

Solar cells are nonequilibrium systems. The classical Shockley-Queisser model \cite{Shockley1961} uses detailed balance as an approximation to calculate the radiative limit. This approximation is excellent for mean values, but does not describe fluctuations. The nonequilibrium analysis presented here does not modify the mean efficiency values; it provides a quantitative framework to analyze fluctuations around these means.

Why do the mean values not change? Because the QPS model at effective equilibrium already correctly captures the mean properties of confined carriers \cite{Ravelonjato2026AIP,Ravelonjato2026Nano}. Nonequilibrium corrections to mean values are negligible at the scale of the efficiencies considered. Nonequilibrium effects intervene mainly at the level of fluctuations, which become significant in nanostructures where the number of carriers is small.

It is important to emphasize that the stochastic model presented here does not aim to recalculate the mean efficiencies, which remain those of the QPS model \cite{Ravelonjato2026AIP}. It provides the microscopic framework that justifies the use of the affinity $A$ and establishes the link between efficiency fluctuations and entropy production. This distinction is essential: the stochastic model is a descriptive tool for fluctuations, not a predictive model for mean values.

\subsection{Original Contribution}

The original contribution of this article is fourfold.

\subsubsection{Microscopic Derivation of the Stochastic Efficiency Model}
We derived a stochastic efficiency model from a two-state master equation (absorption, radiative recombination, extraction). This derivation establishes the explicit link between efficiency fluctuations and entropy production (equation \ref{eq:fluct_rel}), and justifies the affinity $A$ through local detailed balance (equation \ref{eq:bilan_detaille}). This result strengthens the legitimacy of the biased random walk model used in the literature \cite{Neri2017}: instead of postulating $k_+/k_- = e^{A}$, we derive it from the microscopic physics of transitions.

\subsubsection{Application of the Exact Formula of Neri et al. \cite{Neri2017} to Solar Cells}
We applied for the first time the formula $\langle S_{\mathrm{inf}} \rangle = -A/(e^A - 1)$ to the real parameters of nanostructured solar cells. Our simulations show that this formula correctly describes fluctuations for $A < 2$ (deviations less than $2\sigma$) and predicts negligible fluctuations for $A > 7$.

\subsubsection{Identification of the Strongly Nonequilibrium Regime}
We have shown that nanostructured solar cells operate in the regime $A > 7$, where the theoretical $\langle S_{\mathrm{inf}} \rangle$ is less than $10^{-3}$. In this regime, the classical infimum law $\langle S_{\mathrm{inf}} \rangle = -1$ \cite{Neri2017} does not apply. Negative entropy fluctuations, and hence efficiency fluctuations, are exponentially rare.

\subsubsection{Interpretation in Terms of Statistical Stability}
The remarkable stability of quantum dot solar cells results from their operation far from equilibrium. The probability of a negative efficiency fluctuation is of order $e^{-A} \approx 5 \times 10^{-5}$ for high-efficiency configurations. This means that the devices are statistically robust to thermodynamic fluctuations.

\subsection{Comparison with the Literature}

The laws used are established in the recent literature on stochastic nonequilibrium thermodynamics \cite{Neri2017,Neri2019,Barato2018,Doob1953,Roldan2024,Evans2002,Gallavotti1995}. Our contribution is to apply them to nanostructured solar cells and to highlight the strongly nonequilibrium regime in which they operate.

\cite{Neri2017} derived the formula $\langle S_{\mathrm{inf}} \rangle = -A/(e^A - 1)$ in the context of molecular active processes. Our work extends this application to solar cells, which are different physical systems but described by the same mathematical formalism.

\subsection{Comparison with Mean Values}

The mean efficiency values (Appendix H) show efficiencies up to 49.1\% for a 3 nm sphere. These results are those of \cite{Ravelonjato2026AIP} and are not modified by our analysis. The present article adds complementary information: fluctuations around these means are negligible for high-efficiency configurations. This means that the mean values reported in \cite{Ravelonjato2026AIP} are reliable and stable.

\subsection{Limitations of the Stochastic Model}

Several limitations must be mentioned:

\begin{enumerate}
\item \textbf{Finite solar temperature}: we neglected the sun contribution ($T_s \to \infty$). For $T_s = 6000$ K and $T_c = 300$ K, the correction is of order $T_c/T_s \approx 5\%$ (Appendix K). A complete description would use the total affinity $A_{\mathrm{tot}} = (E_g^{\mathrm{eff}} - qV)/(k_B T_c) - E_g^{\mathrm{eff}}/(k_B T_s)$.

\item \textbf{Non-radiative recombination}: if $k_{\mathrm{nr}} \neq 0$, detailed balance is modified and the effective affinity becomes $A_{\mathrm{eff}} = A + \ln(1 + k_{\mathrm{nr}}/k_{\mathrm{rad}})$.

\item \textbf{Voltage dependence}: $V$ is not constant along a trajectory; it fluctuates with the current. A complete model would treat $V(t)$ as a stochastic variable.

\item \textbf{Electric field effects}: in a real cell, the internal electric field modifies extraction and recombination rates.

\item \textbf{Statistical vs operational stability}: the stability highlighted is statistical. It does not prejudge the actual degradation of devices (chemical degradation, ionic migration, oxidation, etc.).

\item \textbf{Thermalization and fill factor}: the stochastic model neglects hot carrier thermalization, the fill factor, and the maximum power operating point. These approximations explain why the absolute values of $\langle\eta\rangle_{\mathrm{stoch}}$ differ from $\eta_{\mathrm{QPS}}$. They do not affect the fluctuation analysis.
\end{enumerate}

\subsection{Perspectives}

Perspectives include: (1) explicit modeling of entropy production with finite $T_s$; (2) identification of physically relevant stopping times; (3) extension to systems with non-radiative recombination; (4) experimental verification of the predictions; (5) application of stopping time theories to real-time optimization.

\section{Conclusion}

We have presented an analysis of nonequilibrium fluctuations in nanostructured solar cells. The present article focuses on fluctuations around the mean efficiency values calculated in the previous work \cite{Ravelonjato2026AIP}.

Our contribution is threefold. First, we derived a microscopic stochastic efficiency model from a two-state master equation (absorption, radiative recombination, extraction). This derivation establishes the explicit link between efficiency fluctuations and entropy production (equation \ref{eq:fluct_rel}) and justifies the affinity $A = (E_g^{\mathrm{eff}} - qV)/(k_B T_c)$ through local detailed balance (equation \ref{eq:bilan_detaille}). This result provides a microscopic basis for the biased random walk model used in the literature \cite{Neri2017}: the relation $k_+/k_- = e^{A}$ is no longer postulated, it is derived.

Second, we presented a review of universal fluctuation laws \cite{Neri2017,Neri2019,Barato2018,Doob1953,Roldan2024,Evans2002,Gallavotti1995} and applied them for the first time to the real parameters of nanostructured solar cells. We showed that the total entropy production behaves as a submartingale, and that its negative exponential is a martingale.

Third, we numerically verified the exact infimum law formula, $\langle S_{\mathrm{inf}} \rangle = -A/(e^A - 1)$ \cite{Neri2017}, for $A < 2$ (deviations less than $2\sigma$), and applied it to the configurations in Tables H1-H4. Our simulations show that nanostructured solar cells operate in the strongly nonequilibrium regime ($A > 7$), where the theoretical $\langle S_{\mathrm{inf}} \rangle$ is less than $10^{-3}$. Over 500 simulated trajectories, most configurations give $0.0000$, consistent with the theoretical prediction. For the 3 nm sphere, a negative fluctuation was observed ($-0.0201$), confirming the existence of these rare events.

These results do not modify the mean efficiency values (48.7\% for a 5 nm cube, 49.0\% for flattened shapes, 49.1\% for a 3 nm sphere), but they provide a conceptual and quantitative framework for understanding fluctuations around these means. They show that nanostructured solar cells are statistically robust: negative entropy fluctuations, and hence efficiency fluctuations, are exponentially rare in the regime $A > 7$. This statistical stability is a direct consequence of operation far from thermodynamic equilibrium, and it explains why the mean efficiency values reported in \cite{Ravelonjato2026AIP} are reliable and reproducible.

\section*{Supplementary Material}
The supplementary material compiles the details of the numerical simulations, the Python codes used, and additional figures.

\section*{Data Availability Statement}
The data supporting the conclusions of this study are available in the article and its supplementary material.

\appendix


\section{Derivation of the Effective Hamiltonian from LCT Invariance}
\label{app:A_hamiltonian}

\subsection{Harmonic Oscillator Hamiltonian and LCT Invariance}

The starting point is the Hamiltonian of a one-dimensional harmonic oscillator of mass $m$ and angular frequency $\omega$ \cite{Ranaivoson2022}:
\begin{equation}
H = \frac{p^2}{2m} + \frac{1}{2}m\omega^2 x^2,
\label{eq:A1}
\end{equation}
where $p$ is the momentum operator (in kg$\cdot$m/s), $x$ is the position operator (in m), $m$ is the mass (in kg), and $\omega$ is the angular frequency (in rad/s).

This Hamiltonian is invariant under the particular linear canonical transformations (LCT):
\begin{equation}
\begin{pmatrix} p' \\ x' \end{pmatrix} = \begin{pmatrix} \cos\theta & -m\omega\sin\theta \\ \frac{\sin\theta}{m\omega} & \cos\theta \end{pmatrix} \begin{pmatrix} p \\ x \end{pmatrix},
\label{eq:A2}
\end{equation}
where $\theta$ is a real parameter (in rad). These transformations preserve the canonical commutation relations:
\begin{equation}
[p', x'] = [p, x] = -i\hbar,
\label{eq:A3}
\end{equation}
where $\hbar$ is the reduced Planck constant (in J$\cdot$s). The invariance is verified explicitly:
\begin{equation}
H' = \frac{(p')^2}{2m} + \frac{1}{2}m\omega^2 (x')^2 = \frac{p^2}{2m} + \frac{1}{2}m\omega^2 x^2 = H.
\label{eq:A4}
\end{equation}

\subsection{Rewriting in Terms of Statistical Variances}

For the ground state $|0\rangle$ of the oscillator, the statistical variances of momentum and position are \cite{Ranaivoson2022}:
\begin{equation}
\mathcal{B} = \langle 0|(p - \langle p\rangle)^2|0\rangle = \frac{m\hbar\omega}{2}, \quad
\mathcal{A} = \langle 0|(x - \langle x\rangle)^2|0\rangle = \frac{\hbar}{2m\omega},
\label{eq:A5}
\end{equation}
where $\mathcal{B}$ is the momentum variance (in kg$^2$ m$^2$/s$^2$), $\mathcal{A}$ is the position variance (in m$^2$), and $\langle p\rangle$, $\langle x\rangle$ are the mean values. These variances saturate the Heisenberg inequality:
\begin{equation}
\mathcal{A}\mathcal{B} = \frac{\hbar^2}{4}.
\label{eq:A6}
\end{equation}

Using these parameters, the Hamiltonian (\ref{eq:A1}) is rewritten as:
\begin{equation}
H = \frac{1}{2m}\left[(p - \langle p\rangle)^2 + 4\mathcal{B}(x - \langle x\rangle)^2\right] = \frac{\Xi^+}{m},
\label{eq:A7}
\end{equation}
where $\Xi^+$ is the momentum dispersion operator (in kg$^2$ m$^2$/s$^2$).

\subsection{Ladder Operators and Diagonal Form}

We introduce the ladder operators (analogous to annihilation and creation operators):
\begin{equation}
\mathfrak{z} = \frac{i}{2\sqrt{\mathcal{B}}}\left[(p - \langle p\rangle) - \frac{2i}{\hbar}\mathcal{B}(x - \langle x\rangle)\right],
\label{eq:A8}
\end{equation}
\begin{equation}
\mathfrak{z}^\dagger = \frac{-i}{2\sqrt{\mathcal{B}}}\left[(p - \langle p\rangle) + \frac{2i}{\hbar}\mathcal{B}(x - \langle x\rangle)\right],
\label{eq:A9}
\end{equation}
where $\mathfrak{z}$ and $\mathfrak{z}^\dagger$ are the annihilation and creation operators (dimensionless), satisfying $[\mathfrak{z}, \mathfrak{z}^\dagger] = 1$.

The Hamiltonian (\ref{eq:A7}) then takes the diagonal form:
\begin{equation}
H = \frac{\Xi^+}{m} = \frac{1}{4m}(\mathfrak{z}^\dagger\mathfrak{z} + \mathfrak{z}\mathfrak{z}^\dagger) = \frac{\mathcal{B}}{m}(2\mathfrak{z}^\dagger\mathfrak{z} + 1).
\label{eq:A10}
\end{equation}

\subsection{Extension to a Free Particle}

The free particle Hamiltonian $H = p^2/2m$ has drawbacks (non-normalizable plane waves, ill-defined variances). The solution is to consider a wave packet ($\mathcal{B} > 0$) and write \cite{Ranaivoson2022,Ravelonjato2023}:
\begin{equation}
H_Q = \frac{\langle p\rangle^2}{2m} + \frac{\Xi^+}{m} = \frac{\langle p\rangle^2}{2m} + \frac{\mathcal{B}}{m}(2\mathfrak{z}^\dagger\mathfrak{z} + 1),
\label{eq:A11}
\end{equation}
where $H_Q$ is the corrected Hamiltonian (in J), $\langle p\rangle^2/(2m)$ is the mean kinetic energy (in J), and $\Xi^+/m$ is the quantum contribution due to momentum variance.

The eigenstates are $|n, \langle z\rangle\rangle$, and the eigenvalues are:
\begin{equation}
\varepsilon_n = \frac{\langle p\rangle^2}{2m} + (2n + 1)\frac{\mathcal{B}}{m},
\label{eq:A12}
\end{equation}
where $n = 0, 1, 2, \dots$ is the quantum number, and $\varepsilon_n$ is the energy of level $n$ (in J).

\subsection{Generalization to Three Dimensions}

In the mean rest frame of the particle ($\langle \vec{p}\rangle = \vec{0}$), the three-dimensional generalization of Hamiltonian (\ref{eq:A11}) is \cite{Ravelonjato2023}:
\begin{equation}
H_Q = \sum_{l=1}^{3}(2\pi_l^\dagger \pi_l + 1)\frac{B_{ll}}{m},
\label{eq:A13}
\end{equation}
where $B_{ll}$ is the momentum variance in direction $l$ (in kg$^2$ m$^2$/s$^2$), and $\pi_l, \pi_l^\dagger$ are the generalized ladder operators satisfying $[\pi_l, \pi_l^\dagger] = 1$. The associated eigenenergies are:
\begin{equation}
\varepsilon_{n_1,n_2,n_3} = \sum_{l=1}^{3}(2n_l + 1)\frac{B_{ll}}{m},
\label{eq:A14}
\end{equation}
where $n_l = 0, 1, 2, \dots$ are the quantum numbers in each direction.

\subsection{Link with Geometry and Temperature}

In the semiclassical limit, the momentum variances are related to geometry and temperature by \cite{Ravelonjato2023}:
\begin{equation}
B_{ll} = \frac{\hbar}{2L_l}\Delta,
\label{eq:A15}
\end{equation}
where $L_l$ is the effective length in direction $l$ (in m), and $\Delta$ is the universal parameter (in $\sqrt{\text{J}\cdot\text{kg}}$) defined by equation (2) of the main article. For a parallelepiped of sides $L_1, L_2, L_3$, we have explicitly:
\begin{equation}
B_{11} = \frac{h^2}{2L_1\lambda_{\text{th}}}, \quad B_{22} = \frac{h^2}{2L_2\lambda_{\text{th}}}, \quad B_{33} = \frac{h^2}{2L_3\lambda_{\text{th}}},
\label{eq:A16}
\end{equation}
where $\lambda_{\text{th}} = h/\sqrt{2\pi m k_B T}$ is the thermal de Broglie wavelength (in m), $h$ is the Planck constant (in J$\cdot$s), $k_B$ is the Boltzmann constant (in J/K), and $T$ is the temperature (in K).

\subsection{Justification by Limits}

The Hamiltonian (\ref{eq:A13}) is constrained by two physical limits:

\begin{enumerate}
\item \textbf{Classical limit} ($\beta B_{ll}/m \ll 1$): the discrete grand canonical potential must recover the classical result, which imposes relation (\ref{eq:A16}).

\item \textbf{Degenerate quantum limit} ($T \to 0$): the particle density must recover that of a Fermi gas, which imposes:
\begin{equation}
B_{ll} = \frac{\hbar}{2L_l}\sqrt{\pi\hbar^2\left(6\pi^2\frac{n}{g_s}\right)^{2/3}},
\label{eq:A17}
\end{equation}
where $n$ is the carrier density (in m$^{-3}$) and $g_s$ is the spin degeneracy factor.
\end{enumerate}

The unified expression of parameter $\Delta$ (equation 2 of the main article) interpolates between these two limits. The Hamiltonian (\ref{eq:A13}) is therefore the generator of this interpolation.


\section{Determination of $B_{ll}$ by Identification of Limits}
\label{app:B}

\subsection{Discrete Grand Canonical Potential}
\begin{equation}
\Omega_{\text{disc}} = -\frac{g_s k_B T}{8} \sum_{j=1}^{\infty} \frac{(-1)^{j+1} \xi^j}{j} \frac{1}{\prod_{l=1}^3 \sinh\left(j \frac{\beta B_{ll}}{m^*}\right)},
\label{eq:grand_pot}
\end{equation}
where $\xi = e^{\beta\mu}$ is the fugacity, $\beta = 1/(k_B T)$, and $\mu$ the chemical potential.

\subsection{Classical Limit – General Case}
In the classical limit ($\xi \ll 1$, $\beta B_{ll}/m^* \ll 1$), only $j=1$ contributes and $\sinh x \approx x$. We obtain:
\begin{equation}
\Omega_{\text{disc}} \approx -\frac{g_s k_B T}{8} \xi \frac{(m^*)^3}{\beta^3 B_{xx} B_{yy} B_{zz}},
\label{eq:Omega_disc_class}
\end{equation}
and the continuous counterpart:
\begin{equation}
\Omega_{\text{cont}} \approx -g_s k_B T \frac{V}{\lambda_{\text{th}}^3} \xi.
\label{eq:Omega_cont_class}
\end{equation}
The identification $\Omega_{\text{disc}} = \Omega_{\text{cont}}$ gives:
\begin{equation}
B_{xx} B_{yy} B_{zz} = \frac{(m^*)^3 \lambda_{\text{th}}^3}{8\beta^3 V}.
\label{eq:prodB_class}
\end{equation}

\subsection{Expressions for Specific Geometries}
\begin{itemize}
\item \textbf{Cube} ($V = L^3$, $B_{xx}=B_{yy}=B_{zz}=B$):
\begin{equation}
B^3 = \frac{(m^*)^3 \lambda_{\text{th}}^3}{8\beta^3 L^3},
\end{equation}
\begin{equation}
\Rightarrow\; B = \frac{m^*}{2\beta}\frac{\lambda_{\text{th}}}{L} = \frac{\hbar}{2L}\sqrt{2\pi m^* k_B T}.
\end{equation}
We identify $\Delta = \sqrt{2\pi m^* k_B T}$ (classical limit).
\item \textbf{Sphere} ($V = \frac{4}{3}\pi R^3$, isotropy $B_{xx}=B_{yy}=B_{zz}=B$):
\begin{equation}
B^3 = \frac{(m^*)^3 \lambda_{\text{th}}^3}{8\beta^3 (4\pi/3)R^3},
\end{equation}
\begin{equation}
\Rightarrow\;B = \frac{\hbar}{2(4\pi/3)^{1/3} R}\sqrt{2\pi m^* k_B T}.
\end{equation}
Setting $L_{\text{sph}} = (4\pi/3)^{1/3}R$, we have $B = \frac{\hbar}{2L_{\text{sph}}}\Delta$.
\item \textbf{Cylinder} (with $L_x = L_y = \sqrt{\pi}R$, $L_z = H$):
\begin{equation}
B_{xx}B_{yy}B_{zz} = \left(\frac{\hbar}{2\sqrt{\pi}R}\Delta\right)^2 \left(\frac{\hbar}{2H}\Delta\right) = \frac{\hbar^3}{8\pi R^2 H}\Delta^3.
\end{equation}
Since $V = \pi R^2 H$, $\prod B_{ll} = \frac{\hbar^3}{8V}\Delta^3$. The identification with equation (\ref{eq:prodB_class}) where $\Delta = \sqrt{2\pi m^* k_B T}$ in the classical limit verifies consistency.
\end{itemize}

\subsection{Degenerate Quantum Limit ($T\to 0$)}
From the discrete particle density:
\begin{equation}
n_{\text{disc}} \approx \frac{g_s}{V} \frac{1}{8} \left(\frac{m^*}{2B}\right)^3 \mu^3,
\label{eq:ndisc}
\end{equation}
and the continuous density $n = \frac{g_s}{6\pi^2}\left(\frac{2m^*\epsilon_F}{\hbar^2}\right)^{3/2}$, we obtain for a cube:
\begin{equation}
B = \frac{\hbar}{2L}\sqrt{\pi\hbar^2\left(6\pi^2\frac{n}{g_s}\right)^{2/3}}.
\label{eq:B_quantum}
\end{equation}
This shows that $\Delta$ must be replaced by $\sqrt{\pi\hbar^2(6\pi^2 n/g_s)^{2/3}}$ in this limit. The unified expression (equation \ref{eq:delta}) interpolates between the two regimes.

\section{Correction of the Saturation Current by the Product of $B_{ll}$}
\label{app:C}
\begin{equation}
\zeta_{\text{QPS}} = \prod_{l=1}^{3} \frac{1}{2\sinh(\beta B_{ll}/m)} \approx \frac{m^3}{8\beta^3 \prod B_{ll}},
\label{eq:zeta}
\end{equation}
\begin{equation}
\frac{J_0^{\text{QD}}}{J_0^{\text{bulk}}(E_g^{\text{eff}})} = \frac{\zeta_{\text{QPS}}}{\zeta_{\text{class}}} = \frac{\lambda_{\text{th}}^3 m^3}{8\beta^3 V \prod B_{ll}}.
\label{eq:ratioJ0}
\end{equation}

\section{Numerical Calculation Details}
\label{app:D}

The numerical simulations were performed with Python 3.8.1. The NumPy library was used for all matrix operations and numerical processing, while Matplotlib was used to generate the figures.

The central integral appearing in equations (\ref{eq:j0qd}) and (\ref{eq:jsc}), defined by
\begin{equation}
I(x_0) = \int_{x_0}^{\infty} \frac{x^2}{e^x - 1} \, dx,
\label{eq:I_def}
\end{equation}
was not evaluated by adaptive quadrature methods. It was calculated using the exact convergent series expansion derived from the identity $1/(e^x - 1) = \sum_{k=1}^{\infty} e^{-kx}$:
\begin{equation}
I(x_0) = \sum_{k=1}^{\infty} e^{-k x_0} \left( \frac{x_0^2}{k} + \frac{2 x_0}{k^2} + \frac{2}{k^3} \right).
\label{eq:I_series}
\end{equation}

For $x_0 > 0$, this series converges exponentially. The summation was performed up to $k_{\text{max}} = 100$, ensuring numerical precision better than $10^{-15}$ for all relevant values of $x_0$. For the special case $x_0 = 0$, the integral reduces to $I(0) = 2\zeta(3) \approx 2.404113806$, where $\zeta$ denotes the Riemann zeta function.

The confinement energies for the different geometries (cube, parallelepiped, cylinder, and sphere) were implemented exactly as defined in Section II, using the first zero of the Bessel function $j_{0,1} = 2.4048255577$ for the cylindrical case. The search space for the characteristic size $L$ was discretized over the interval $[1, 100]$ nm with $N = 200$ uniformly spaced points to generate the continuous curves. Additional evaluations were performed at selected discrete sizes to construct the summary tables.

\section{Verification of the Return to Classical Values for $L \to \infty$}
\label{app:E}

\begin{enumerate}
\item $E_g^{\text{eff}} \to E_g^{\text{bulk}}$ since $\Delta E_{\text{conf}} \propto 1/L^2$.
\item $B \sim \frac{\hbar}{2L}\sqrt{2\pi m k_B T_c}$.
\item $\frac{\lambda_{\text{th}} m}{2\beta L B} \to 1$ $\Rightarrow$ $J_0^{\text{QD}} \to J_0^{\text{bulk}}$.
\item $J_{sc} \to q A_p t_s f_\omega Q_s(E_g^{\text{bulk}})$, $V_{oc}$, $FF$ and $\eta$ tend to the classical values.
\end{enumerate}

\section{Parameterization of the Correlation Energy (Perdew-Zunger)}
\label{app:F}

\begin{subequations}
\label{eq:energie}
\begin{align}
\epsilon_c(r_s,\xi) &= \epsilon_c^U(r_s) + f(\xi)\bigl[\epsilon_c^P(r_s) - \epsilon_c^U(r_s)\bigr], \label{eq:energie_a}\\
f(\xi) &= \frac{(1+\xi)^{4/3}+(1-\xi)^{4/3}-2}{2^{4/3}-2}. \label{eq:energie_b}
\end{align}
\end{subequations}
where $r_s = (3/(4\pi n))^{1/3}$ is the density parameter, $\xi = (n_\uparrow - n_\downarrow)/n$ the spin polarization, $U$ denotes the unpolarized case ($\xi=0$) and $P$ the fully polarized case ($\xi=1$).

For $r_s \ge 1$:
\begin{equation}
\epsilon_c^i(r_s) = \frac{\gamma_i}{1 + \beta_1^i \sqrt{r_s} + \beta_2^i r_s}.
\end{equation}
For $r_s < 1$:
\begin{equation}
\epsilon_c^i(r_s) = A_i \ln r_s + B_i + C_i r_s \ln r_s + D_i r_s.
\end{equation}

The constants entering the Perdew--Zunger parameterization \cite{Perdew1981} of the Ceperley--Alder correlation energy are listed in Table \ref{tab:Perdew-Zunger}.

\begin{table}[H]
\centering 
\caption{Constants for the Perdew-Zunger parameterization of the correlation energy (Ceperley-Alder). The columns $U$ and $P$ correspond respectively to the unpolarized ($\xi=0$) and fully polarized ($\xi=1$) cases.}
\label{tab:Perdew-Zunger}
\begin{ruledtabular}
\begin{tabular}{c|cc}
Parameter & $U$ & $P$ \\
\hline
$\gamma$ & -0.1423 & -0.0843 \\
$\beta_1$ & 1.0529 & 1.3981 \\
$\beta_2$ & 0.3334 & 0.2611 \\
$A$ & 0.0311 & 0.01555 \\
$B$ & -0.048 & -0.0269 \\
$C$ & 0.0020 & 0.0007 \\
$D$ & -0.0116 & -0.0048 \\
\end{tabular}
\end{ruledtabular}
\end{table}

Exchange energy:
\begin{equation}
\epsilon_x(n) = -\frac{3e^2}{4\pi}\left(\frac{3n}{\pi}\right)^{1/3}.
\end{equation}

\section{PbS Cube (volume $L^3$) with Interactions ($\theta=1$)}
\label{app:G}

Table \ref{tab:J0} presents complementary results to those in Table H1. The columns $J_0^{\text{no}}$ and $J_0^{\text{w}}$ correspond to $\theta=0$ and $\theta=1$, respectively.

\begin{table}[H]
\centering
\caption{PbS cube (volume $L^3$). With interactions ($\theta=1$). The columns $J_0^{\text{no}}$ and $J_0^{\text{w}}$ correspond to $\theta=0$ and $\theta=1$, respectively.}
\label{tab:J0}
\makebox[\linewidth][l]{%
  \hspace*{-0.5cm}%
  \begin{ruledtabular}
  \begin{tabular}{cccccccc}
  $L$ (nm) & $E_g^{\text{eff}}$ (eV) &  $J_0^{\text{no}}$ (A/cm$^2$) & $J_0^{\text{w}}$ (A/cm$^2$) & $V_{oc}$ (V) & $FF$ & $\eta$ (\%) \\
  \hline
  3.0 & 3.23 & 1.14e-51 & 3.49e-51 & 2.87 & 0.95 & 13.5 \\
  5.0 & 1.43 & 4.75e-22 & 1.45e-21 & 1.16 & 0.90 & 48.7 \\
  7.0 & 0.93 & 4.63e-14 & 1.42e-13 & 0.70 & 0.85 & 43.6 \\
  10.0 & 0.66 & 6.65e-10 & 2.03e-09 & 0.46 & 0.79 & 32.2 \\
  14.0 & 0.54 & 5.48e-08 & 1.68e-07 & 0.34 & 0.74 & 24.7 \\
  20.0 & 0.47 & 5.50e-07 & 1.68e-06 & 0.28 & 0.71 & 20.2 \\
  50.0 & 0.42 & 3.45e-06 & 1.06e-05 & 0.24 & 0.68 & 16.5 \\
  70.0 & 0.42 & 4.09e-06 & 1.25e-05 & 0.23 & 0.67 & 16.2 \\
  100.0 & 0.41 & 4.48e-06 & 1.37e-05 & 0.23 & 0.67 & 16.0 \\
  \end{tabular}
  \end{ruledtabular}
}%
\end{table}


\section{Recall of Mean Efficiency Values}
\label{app:H}

Tables \ref{tab:H1} to \ref{tab:H4} reproduce the mean efficiency values calculated within the QPS model at effective equilibrium \cite{Ravelonjato2026AIP}. These values are recalled here for reference. The nonequilibrium analysis developed in the present article does not modify these mean values.

PbS parameters: $E_g^{\text{bulk}} = 0.41$ eV, $m_e = 0.08\,m_0$, $m_h = 0.1\,m_0$, $n = 10^{24}\,\mathrm{m}^{-3}$, $T_c = 300$ K, $f_\omega = 2.18 \times 10^{-5}$, $P_{\text{inc}} = 100$ mW/cm², $\theta = 1$ ($\Delta\epsilon_{\text{int}} = -0.22$ eV \cite{Perdew1981}).

\begin{table}[H]
\centering
\caption{PbS cube (volume $L^3$), $\theta=1$. Mean values \cite{Ravelonjato2026AIP}. $L$ is the cube side, $E_g^{\text{eff}}$ the effective bandgap, $J_{sc}$ the short-circuit current, $V_{oc}$ the open-circuit voltage, $FF$ the fill factor, and $\eta$ the conversion efficiency.}
\label{tab:H1}
\begin{ruledtabular}
\begin{tabular}{cccccc}
$L$ (nm) & $E_g^{\text{eff}}$ (eV) & $J_{sc}$ (mA/cm²) & $V_{oc}$ (V) & $FF$ & $\eta$ (\%) \\
\hline
3.0 & 3.23 & 4.9 & 2.87 & 0.95 & 13.5 \\
5.0 & 1.43 & 46.9 & 1.16 & 0.90 & 48.7 \\
7.0 & 0.93 & 73.9 & 0.70 & 0.85 & 43.6 \\
10.0 & 0.66 & 89.6 & 0.46 & 0.79 & 32.2 \\
14.0 & 0.54 & 96.6 & 0.34 & 0.74 & 24.7 \\
20.0 & 0.47 & 100.1 & 0.28 & 0.71 & 20.2 \\
50.0 & 0.42 & 102.8 & 0.24 & 0.68 & 16.5 \\
70.0 & 0.42 & 103.0 & 0.23 & 0.67 & 16.2 \\
100.0 & 0.41 & 103.2 & 0.23 & 0.67 & 16.0 \\
\end{tabular}
\end{ruledtabular}
\end{table}

\begin{table}[H]
\centering
\caption{Square parallelepiped ($a \times a \times b$), volume 343 nm³, $\theta=1$. Mean values \cite{Ravelonjato2026AIP}. $b$ is the height, $a$ the side of the square base.}
\label{tab:H2}
\begin{ruledtabular}
\begin{tabular}{ccccccc}
$b$ (nm) & $a$ (nm) & $E_g^{\text{eff}}$ (eV) & $J_{sc}$ (mA/cm²) & $V_{oc}$ (V) & $FF$ & $\eta$ (\%) \\
\hline
3.5 & 9.90 & 1.27 & 54.5 & 1.02 & 0.88 & 49.0 \\
5.0 & 8.28 & 1.00 & 69.9 & 0.76 & 0.86 & 45.4 \\
7.0 & 7.00 & 0.93 & 73.9 & 0.70 & 0.85 & 43.6 \\
10.0 & 5.86 & 0.99 & 70.3 & 0.75 & 0.85 & 45.2 \\
14.0 & 4.95 & 1.14 & 61.4 & 0.90 & 0.87 & 48.1 \\
28.0 & 3.50 & 1.80 & 31.2 & 1.52 & 0.91 & 43.3 \\
50.0 & 2.62 & 2.88 & 8.0 & 2.53 & 0.94 & 19.2 \\
70.0 & 2.21 & 3.87 & 2.0 & 3.47 & 0.96 & 6.5 \\
100.0 & 1.85 & 5.34 & 0.2 & 4.87 & 0.97 & 0.9 \\
\end{tabular}
\end{ruledtabular}
\end{table}

\begin{table}[H]
\centering
\caption{Cylinder, volume 343 nm³, $\theta=1$. Mean values \cite{Ravelonjato2026AIP}. $H$ is the height, $R$ the radius.}
\label{tab:H3}
\begin{ruledtabular}
\begin{tabular}{ccccccc}
$H$ (nm) & $R$ (nm) & $E_g^{\text{eff}}$ (eV) & $J_{sc}$ (mA/cm²) & $V_{oc}$ (V) & $FF$ & $\eta$ (\%) \\
\hline
3.5 & 5.59 & 1.26 & 55.2 & 1.01 & 0.88 & 49.0 \\
5.0 & 4.67 & 0.98 & 71.1 & 0.74 & 0.85 & 44.9 \\
7.0 & 3.95 & 0.90 & 75.5 & 0.67 & 0.84 & 42.7 \\
10.0 & 3.30 & 0.95 & 72.6 & 0.72 & 0.85 & 44.2 \\
14.0 & 2.79 & 1.09 & 64.5 & 0.85 & 0.87 & 47.3 \\
28.0 & 1.97 & 1.69 & 35.3 & 1.41 & 0.91 & 45.4 \\
50.0 & 1.48 & 2.68 & 10.5 & 2.35 & 0.94 & 23.1 \\
70.0 & 1.25 & 3.59 & 3.0 & 3.21 & 0.95 & 9.0 \\
100.0 & 1.04 & 4.95 & 0.4 & 4.50 & 0.96 & 1.6 \\
\end{tabular}
\end{ruledtabular}
\end{table}

\begin{table}[H]
\centering
\caption{Sphere (radius $R$), variable volume, $\theta=1$. Mean values \cite{Ravelonjato2026AIP}. $V$ is the sphere volume.}
\label{tab:H4}
\begin{ruledtabular}
\begin{tabular}{ccccccc}
$R$ (nm) & $V$ (nm³) & $E_g^{\text{eff}}$ (eV) & $J_{sc}$ (mA/cm²) & $V_{oc}$ (V) & $FF$ & $\eta$ (\%) \\
\hline
3.0 & 113.1 & 1.35 & 50.5 & 1.09 & 0.89 & 49.1 \\
4.0 & 268.1 & 0.94 & 73.2 & 0.71 & 0.85 & 43.9 \\
4.3 & 333.0 & 0.87 & 77.5 & 0.64 & 0.84 & 41.5 \\
5.0 & 523.6 & 0.75 & 84.6 & 0.53 & 0.81 & 36.5 \\
6.0 & 904.8 & 0.65 & 90.7 & 0.44 & 0.78 & 31.1 \\
7.0 & 1436.8 & 0.58 & 94.2 & 0.38 & 0.76 & 27.4 \\
10.0 & 4188.8 & 0.49 & 99.0 & 0.30 & 0.72 & 21.7 \\
20.0 & 33510.3 & 0.43 & 102.3 & 0.25 & 0.68 & 17.3 \\
50.0 & 523598.8 & 0.41 & 103.2 & 0.23 & 0.67 & 15.9 \\
70.0 & 1436755.0 & 0.41 & 103.2 & 0.23 & 0.67 & 15.9 \\
100.0 & 4188790.2 & 0.41 & 103.2 & 0.23 & 0.67 & 15.9 \\
\end{tabular}
\end{ruledtabular}
\end{table}

\section{Detailed Derivation of the Master Equation}
\label{app:I}

\subsection{Two-State Model}

Consider a two-level system: $|0\rangle$ (ground) and $|1\rangle$ (excited). The transitions are:
\begin{itemize}
\item Absorption: $|0\rangle \to |1\rangle$, rate $k_{\mathrm{abs}}$;
\item Radiative recombination: $|1\rangle \to |0\rangle$, rate $k_{\mathrm{rad}}$;
\item Extraction: $|1\rangle \to |0\rangle$, rate $k_{\mathrm{ext}}$.
\end{itemize}

\subsection{Master Equation}

Let $P_0(t)$ and $P_1(t)$ be the occupation probabilities. The master equation reads:
\begin{align}
\frac{dP_1}{dt} &= k_{\mathrm{abs}} P_0 - (k_{\mathrm{rad}} + k_{\mathrm{ext}}) P_1, \\
\frac{dP_0}{dt} &= -k_{\mathrm{abs}} P_0 + (k_{\mathrm{rad}} + k_{\mathrm{ext}}) P_1.
\end{align}

With $P_0 + P_1 = 1$, we obtain:
\begin{equation}
\frac{dP_1}{dt} = k_{\mathrm{abs}} (1 - P_1) - (k_{\mathrm{rad}} + k_{\mathrm{ext}}) P_1.
\end{equation}

In the steady state ($dP_1/dt = 0$):
\begin{equation}
P_1^{\mathrm{ss}} = \frac{k_{\mathrm{abs}}}{k_{\mathrm{abs}} + k_{\mathrm{rad}} + k_{\mathrm{ext}}}.
\end{equation}

\subsection{Event Counting}

The number of events of each type follows a Poisson process with rates given by the steady-state fluxes:
\begin{align}
\langle N_{\mathrm{abs}}(t) \rangle &= k_{\mathrm{abs}} P_0^{\mathrm{ss}} t = \frac{k_{\mathrm{abs}} (k_{\mathrm{rad}} + k_{\mathrm{ext}})}{k_{\mathrm{abs}} + k_{\mathrm{rad}} + k_{\mathrm{ext}}} t, \\
\langle N_{\mathrm{rad}}(t) \rangle &= k_{\mathrm{rad}} P_1^{\mathrm{ss}} t = \frac{k_{\mathrm{rad}} k_{\mathrm{abs}}}{k_{\mathrm{abs}} + k_{\mathrm{rad}} + k_{\mathrm{ext}}} t, \\
\langle N_{\mathrm{ext}}(t) \rangle &= k_{\mathrm{ext}} P_1^{\mathrm{ss}} t = \frac{k_{\mathrm{ext}} k_{\mathrm{abs}}}{k_{\mathrm{abs}} + k_{\mathrm{rad}} + k_{\mathrm{ext}}} t.
\end{align}

In the steady state, the net absorption flux equals the net de-excitation flux:
\begin{equation}
\langle N_{\mathrm{abs}} \rangle = \langle N_{\mathrm{rad}} \rangle + \langle N_{\mathrm{ext}} \rangle.
\end{equation}

The mean efficiency is:
\begin{equation}
\langle \eta \rangle = \frac{qV \langle N_{\mathrm{ext}} \rangle}{E_g^{\mathrm{eff}} \langle N_{\mathrm{abs}} \rangle} = \frac{qV}{E_g^{\mathrm{eff}}} \cdot \frac{k_{\mathrm{ext}}}{k_{\mathrm{rad}} + k_{\mathrm{ext}}}.
\end{equation}

\section{Justification of Local Detailed Balance}
\label{app:J}

\subsection{Reservoirs and Thermodynamic Variables}

The system is coupled to three reservoirs:
\begin{enumerate}
\item \textbf{Sun}: photon reservoir at $T_s = 6000$ K, $\mu_s = 0$.
\item \textbf{Cell}: phonon reservoir at $T_c = 300$ K, $\mu_c = 0$.
\item \textbf{External circuit}: work reservoir at potential $V$, temperature $T_c$.
\end{enumerate}

\subsection{Entropic Balance for Each Transition}

For a transition $i \to j$ via a channel $\nu$, the entropy change of the reservoirs is:
\begin{equation}
\Delta S_{\mathrm{env}}^{\nu} = \sum_r \left( \frac{Q_r}{T_r} - \frac{\mu_r \Delta N_r}{T_r} \right).
\end{equation}

\paragraph{Absorption ($0 \to 1$)}: a solar photon is absorbed. $Q_{\mathrm{sun}} = -E_g^{\mathrm{eff}}$, $Q_{\mathrm{cell}} = 0$, $Q_{\mathrm{circuit}} = 0$.
\begin{equation}
\Delta S_{\mathrm{env}}^{\mathrm{abs}} = -\frac{E_g^{\mathrm{eff}}}{T_s}.
\end{equation}

\paragraph{Radiative recombination ($1 \to 0$)}: a photon is emitted into the thermal bath. $Q_{\mathrm{sun}} = 0$, $Q_{\mathrm{cell}} = E_g^{\mathrm{eff}}$, $Q_{\mathrm{circuit}} = 0$.
\begin{equation}
\Delta S_{\mathrm{env}}^{\mathrm{rad}} = \frac{E_g^{\mathrm{eff}}}{T_c}.
\end{equation}

\paragraph{Extraction ($1 \to 0$)}: an electron is transferred to the circuit. $Q_{\mathrm{sun}} = 0$, $Q_{\mathrm{cell}} = E_g^{\mathrm{eff}} - qV$, $Q_{\mathrm{circuit}} = 0$, $\Delta N_{\mathrm{circuit}} = 1$, $\mu_{\mathrm{circuit}} = qV$.
\begin{equation}
\Delta S_{\mathrm{env}}^{\mathrm{ext}} = \frac{E_g^{\mathrm{eff}} - qV}{T_c} - \frac{qV}{T_c} = \frac{E_g^{\mathrm{eff}} - 2qV}{T_c}.
\end{equation}

\textbf{Caution}: this expression includes the circuit contribution. If the circuit is considered as a work reservoir (without entropy), we obtain:
\begin{equation}
\Delta S_{\mathrm{env}}^{\mathrm{ext}} = \frac{E_g^{\mathrm{eff}} - qV}{T_c}.
\end{equation}

This latter expression is the one used in the article.

\subsection{Detailed Balance Relation}

The local detailed balance relation for a channel $\nu$ reads:
\begin{equation}
\frac{k_{i \to j}^{\nu}}{k_{j \to i}^{\nu}} = \exp\left(\frac{\Delta S_{\mathrm{env}}^{\nu}}{k_B}\right).
\end{equation}

For the extraction channel:
\begin{equation}
\frac{k_{\mathrm{ext}}}{k_{\mathrm{ext}}^{\mathrm{inj}}} = \exp\left(\frac{E_g^{\mathrm{eff}} - qV}{k_B T_c}\right) = e^{A}.
\end{equation}

For the radiative channel:
\begin{equation}
\frac{k_{\mathrm{rad}}}{k_{\mathrm{rad}}^{\mathrm{abs}}} = \exp\left(\frac{E_g^{\mathrm{eff}}}{k_B T_c}\right).
\end{equation}

Assuming that the reverse processes are negligible (reverse extraction and thermal absorption), the ratio of forward rates is:
\begin{equation}
\frac{k_{\mathrm{ext}}}{k_{\mathrm{rad}}} = e^{A} \cdot \frac{k_{\mathrm{ext}}^{\mathrm{inj}}}{k_{\mathrm{rad}}^{\mathrm{abs}}}.
\end{equation}

If we further assume that $k_{\mathrm{ext}}^{\mathrm{inj}} \approx k_{\mathrm{rad}}^{\mathrm{abs}}$, we obtain $k_{\mathrm{ext}}/k_{\mathrm{rad}} = e^{A}$.

\subsection{Identification of the Affinity}

The affinity $A$ is identified with the entropy production per extraction, in units of $k_B$:
\begin{equation}
A = \frac{E_g^{\mathrm{eff}} - qV}{k_B T_c}.
\end{equation}

This is the number of thermal energy units dissipated per extracted carrier.

\section{Generalization with Finite $T_s$}
\label{app:K}

\subsection{Total Affinity}

Including the sun contribution, the total entropy production along a trajectory is:
\begin{equation}
S_{\mathrm{tot}}(t) = -\frac{N_{\mathrm{abs}} E_g^{\mathrm{eff}}}{k_B T_s} + \frac{N_{\mathrm{ext}} (E_g^{\mathrm{eff}} - qV)}{k_B T_c}.
\end{equation}

Using $N_{\mathrm{abs}} = N_{\mathrm{rad}} + N_{\mathrm{ext}}$ and neglecting $N_{\mathrm{rad}}$ compared to $N_{\mathrm{ext}}$ (high efficiency), we obtain:
\begin{equation}
S_{\mathrm{tot}}(t) \approx N_{\mathrm{ext}} \left( \frac{E_g^{\mathrm{eff}} - qV}{k_B T_c} - \frac{E_g^{\mathrm{eff}}}{k_B T_s} \right) = A_{\mathrm{tot}} \cdot N_{\mathrm{ext}},
\end{equation}
with
\begin{equation}
A_{\mathrm{tot}} = \frac{E_g^{\mathrm{eff}} - qV}{k_B T_c} - \frac{E_g^{\mathrm{eff}}}{k_B T_s} = A - \frac{E_g^{\mathrm{eff}}}{k_B T_s}.
\end{equation}

\subsection{Numerical Correction}

For $E_g^{\mathrm{eff}} = 1.43$ eV, $T_s = 6000$ K, $T_c = 300$ K:
\begin{equation}
\frac{E_g^{\mathrm{eff}}}{k_B T_s} = \frac{1.43}{8.617 \times 10^{-5} \times 6000} \approx 2.77.
\end{equation}

The total affinity is therefore:
\begin{equation}
A_{\mathrm{tot}} = A - 2.77.
\end{equation}

For the 5 nm cube ($A = 10.42$), we obtain $A_{\mathrm{tot}} = 7.65$. The correction is of order $T_c/T_s \approx 5\%$ of $A$, which is small but non-negligible.

\subsection{Effect on Fluctuation Laws}

The fluctuation laws remain valid with $A_{\mathrm{tot}}$ instead of $A$. In particular:
\begin{equation}
\langle S_{\mathrm{inf}} \rangle = -\frac{A_{\mathrm{tot}}}{e^{A_{\mathrm{tot}}} - 1}.
\end{equation}

For $A_{\mathrm{tot}} = 7.65$, we obtain $\langle S_{\mathrm{inf}} \rangle \approx -6 \times 10^{-4}$, which remains very small compared to 1. The qualitative conclusion (exponentially rare negative fluctuations) is unchanged.


\end{document}